\documentclass[twocolumn,showpacs,preprintnumbers,prb,amsmath,amssymb]{revtex4}
\usepackage{graphicx}% Include figure files
\usepackage{dcolumn}% Align table columns on decimal point
\usepackage{bm}% bold math

\begin{document}

%\preprint{APS/123-QED}

\title{Polarization properties of single photons emitted by nitrogen-vacancy defect in diamond at low temperature}% Force line breaks with \\
\author{F.~Kaiser$^{1}$, V. Jacques$^{1,\ast}$, A. Batalov$^{1}$, P. Siyushev$^{1}$, F. Jelezko$^{1}$ and J. Wrachtrup$^{1}$}

%\email{v.jacques@physik.uni-stuttgart.de}

\affiliation{$^{1}$3. Physikalisches Institut, Universit$\ddot{a}$t Stuttgart, 70550 Stuttgart, Germany}

\altaffiliation[Electronic address: ]{v.jacques@physik.uni-stuttgart.de}

\date{\today}

\begin{abstract}
In this report, the polarization properties of the photoluminescence emitted by single nitrogen-vacancy (NV) color centers in diamond are investigated using resonant excitation at cryogenic temperature. We first underline that the two excited-state orbital branches are associated with two orthogonal transition dipoles. Using selective excitation of one dipole, we then show that the photoluminescence is partially unpolarized owing to fast relaxation between the two orbitals induced by the thermal bath. This result might be important in the context of the realization of indistinguishable single photons using NV defect in diamond.
\end{abstract}

\maketitle

\begin{figure}[t]
\centerline{\includegraphics[width=8cm]{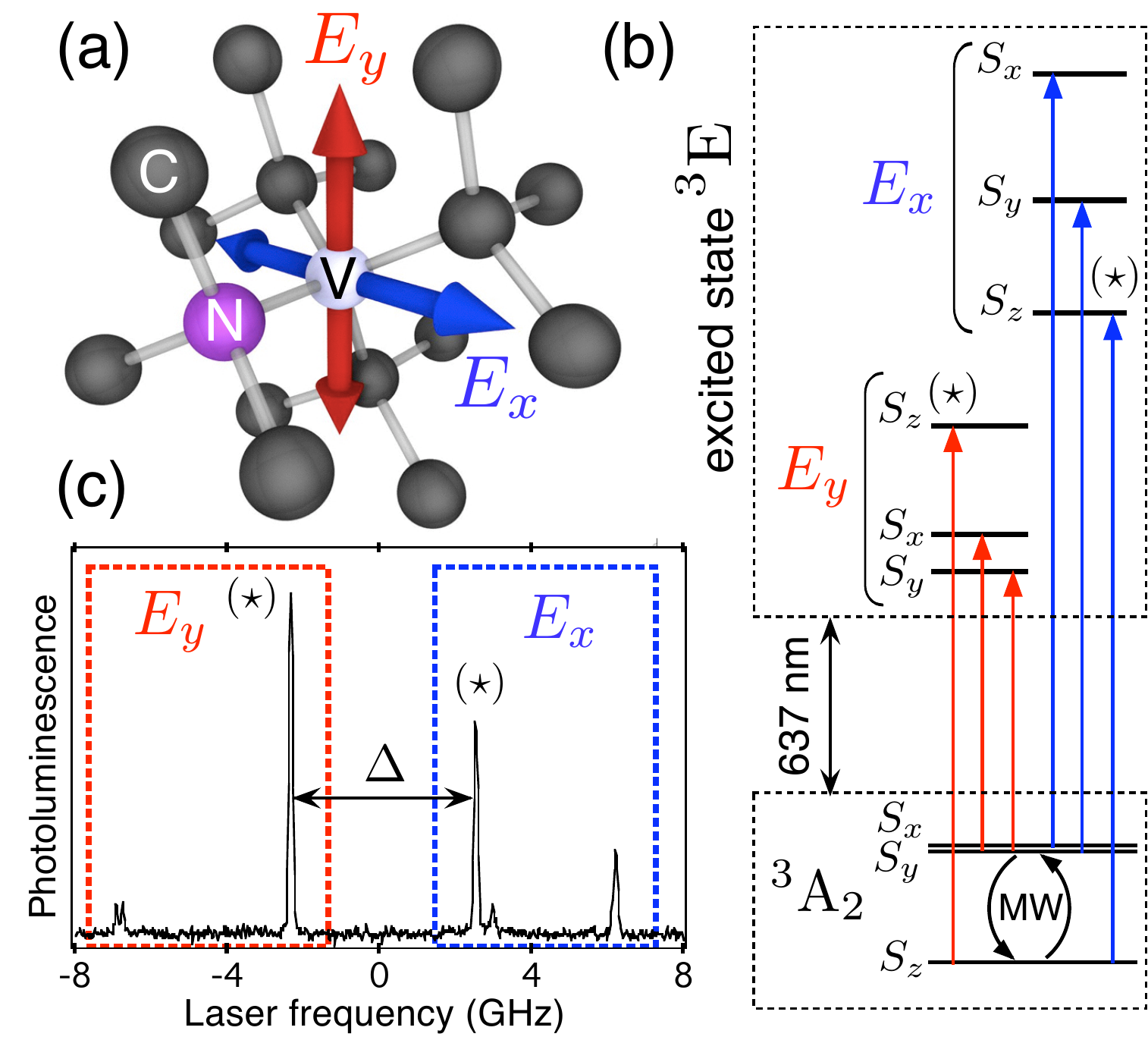}}
\caption{(color online). (a)-Atomic structure of the nitrogen-vacancy (NV) defect in diamond indicating two orthogonal dipoles associated with the excited-state orbital branches $E_{x}$ and $E_{y}$, in the plane perpendicular to the symmetry axis $z$. (b)-Energy-level diagram, showing the spin triplet ground state $^{3}$A$_{2}$ and the spin triplet excited-state $^{3}$E, split into two orbital branches $E_{x}$ and $E_{y}$ by local strain. (c)-Excitation spectrum of a single NV defect. Resonances marked with the symbol $(\star)$ correspond to transitions linking $S_{z}$ spin sublevels.}
\label{Fig1}
\end{figure}

\indent The nitrogen-vacancy (NV) color center in diamond is an outstanding candidate for solid-state quantum information processing because its spin state can be coherently manipulated with high fidelity owing to long coherence time, even at room temperature~\cite{Fedor_revue}. The electron spin of a single NV defect behaves as an ultrasensitive magnetometer at the nanoscale~\cite{Maze_Nature2008,Gupi_Nature2008}, allowing to optically detect magnetic dipolar coupling with neighboring single electron and nuclear spins of the diamond crystalline matrix~\cite{Lukin_Science2006,Hanson_Science2008}. Such magnetic coupling has recently been used as a resource for realizing quantum registers~\cite{Lukin_Science2007} and multipartite entanglement among single spins in diamond~\cite{Neumann_Science2008}. The maximum distance for which magnetic dipolar coupling can be detected depends on the NV defect electron spin coherence time and on the spin-spin interaction strength. Although coherence times up to several milliseconds have been measured in ultra-pure diamond samples~\cite{Gupi_NatMat2009}, a few tens of nanometers is the limit distance to detect magnetic coupling between two independent NV defect and thus to create entanglement between local quantum registers. \\
\indent A way to achieve entanglement between NV defects over macroscopic distances relies on coupling between spin states (stationary qubit) and photons (flying qubit)~\cite{Barrett_PRA2005,Moehring_Nature2007,Olmschenk_Science2009,Lukin_PRL2006}. Performing such long-distance entanglement protocols requires the emission of indistinguishable single photons from each single emitter, {\it i.e.} Fourier transform relation between their spectral and temporal profiles, same spatial mode and identical polarizations. Although Fourier-transform emission from single NV defect in diamond has recently been reported at cryogenic temperatures~\cite{Tamarat_PRL2006,Batalov_PRL2008}, the polarization properties of the photoluminescence (PL) have not been investigated in details so far. Such knowledge is however of crucial importance in order to achieve emission of indistinguishable single photons.\\
\indent In this letter, we study the polarization properties of the PL emitted by single NV defect in diamond. Using resonant excitation at low temperature, we first show that the two excited-state orbitals are associated with two orthogonal transition dipoles. The polarization properties of emitted photons are then analysed using selective excitation of a given transition dipole. We find that the PL is partially unpolarized owing to relaxation between the two orbitals induced by the thermal bath. \\
\indent The NV color center in diamond consists of a substitutional nitrogen atom (N) associated with a vacancy (V) at an adjacent lattice site, giving a defect with $\rm C_{3v}$ symmetry (Fig.~\ref{Fig1}(a)). For the negatively charged NV color center addressed in this study, the ground state is a spin triplet state $^{3} \rm A_{2}$ in which the degeneracy is lifted by spin-spin interaction into a spin doublet $S_{x},S_{y}$ and a spin singlet $S_{z}$, where $z$ is the NV symmetry axis~\cite{Manson_PRB2006,Gali_PRB2008,Newton_PRB2009}. The excited state $^{3} \rm E$ is also a spin triplet, associated with a broadband PL emission with zero phonon line (ZPL) around $637$ nm ($1.945$ eV), which allows optical detection of single NV defects using confocal microscopy~\cite{Gruber_Science1997,Kurtsiefer_PRL2000,Beveratos_OptLett2000}. Besides, the $^{3} \rm E$ excited state is an orbital doublet, in which the degeneracy is lifted by non-axial local strain into two orbitals, $E_{x}$ and $E_{y}$, each orbital branch being formed by three spin states $S_{x}$,$S_{y}$ and $S_{z}$ (see Fig.~\ref{Fig1}(b))~\cite{Tamarat_NJP2008,Batalov_PRL2009}. Optical transitions $^{3} \rm A_{2}\rightarrow ^{3}$E are spin-conserving. Furthermore, owing to $\rm C_{3v}$ symmetry of the defect, electric dipole transitions are allowed for dipoles in the plane perpendicular to the symmetry axis of the NV defect ($z$ axis). Thereby, the excited-state orbital branches $E_{x}$ and $E_{y}$ are expected to be associated with two orthogonal dipoles located in the $(x,y)$ plane (see Fig.~\ref{Fig1}(a))~\cite{Davies,Epstein_NatPhys2005,Santori_PRB2007}.\\
\indent Recently, resonant optical excitation of single NV defect at low temperature has shown that the optical transition $\left|S_{z} \right\rangle \rightarrow \left|E_{x},S_{z}\right\rangle$ is the only cycling transition~\cite{Tamarat_NJP2008,Batalov_PRL2009}. All other transitions are not cycling as spin-flips can likely occur, either by inter-system crossing to the metastable state, or through a mixing of the excited-state spin sublevels by non-axial spin orbit interaction and local strain. Experimental observation of the excited-state orbital doublet then requires the use of a microwave (MW) excitation resonant with the ground state transition, in order to maintain a time-averaged population within each of the ground state spin sublevels~\cite{Tamarat_NJP2008,Batalov_PRL2009}.\\
\indent  In the following, we investigate single NV defects in a [111]-oriented type IIa natural diamond, where the nitrogen concentration is below 1 ppm. Among the four possible orientations of NV defects in the diamond matrix, along [$111$], [$\bar{1}\bar{1}1$], [$1\bar{1}1$], or [$\bar{1}11$] direction, we restrict the study to [$111$]-oriented defects, for which the quantization axis $z$ is normal to the sample surface. For such an orientation, the two transition dipoles $E_{x}$ and $E_{y}$ are therefore in the diamond surface plane, which is the best configuration for studying polarization properties of single NV defects emission. As the PL intensity is stronger for [$111$]-oriented defects compared to the three other orientations, it is possible to distinguish optically  [$111$]-oriented NV defects in the sample~\cite{Santori_PRB2007}. Note that the NV defect orientation has also been checked by recording electron spin resonance (ESR) spectra with a magnetic field applied along the [$111$] axis. As the angle between the magnetic field and the NV defect quantization axis changes for different orientations, a measurement of Zeeman splitting in ESR spectrum allows to distinguish NV defects with [111] orientation~\cite{Santori_PRB2007}.\\
\indent Single NV defects are imaged using confocal microscopy at cryogenic temperatures ($T\approx 4$ K). A tunable laser diode is used to excite resonantly NV centers on their ZPL while the red-shifted PL between 650 nm and 750 nm is detected in a confocal arrangement and used to monitor excitation spectra by sweeping the laser diode frequency. In addition, MW resonant with the ground state spin transition are applied via a $20 \ \mu$m diameter copper microwire located close to the NV defects, in order to observe the full structure of the excited-state. A typical excitation spectrum of a single NV defect is depicted in Fig.~\ref{Fig1}(c). Three optical transitions are observed for each excited-state orbital $E_{x}$ and $E_{y}$, corresponding to transitions between identical spin sublevels. In the following, we focus on the transitions linking $S_{z}$ spin sublevels for each orbital branch (see $(\star)$ symbols in Fig.~\ref{Fig1}(c)), which are the most efficient transitions owing to the low shelving rate to the metastable state from $S_{z}$ spin sublevels~\cite{Manson_PRB2006}. 
\begin{figure}[b]
\centerline{\includegraphics[width=8.5cm]{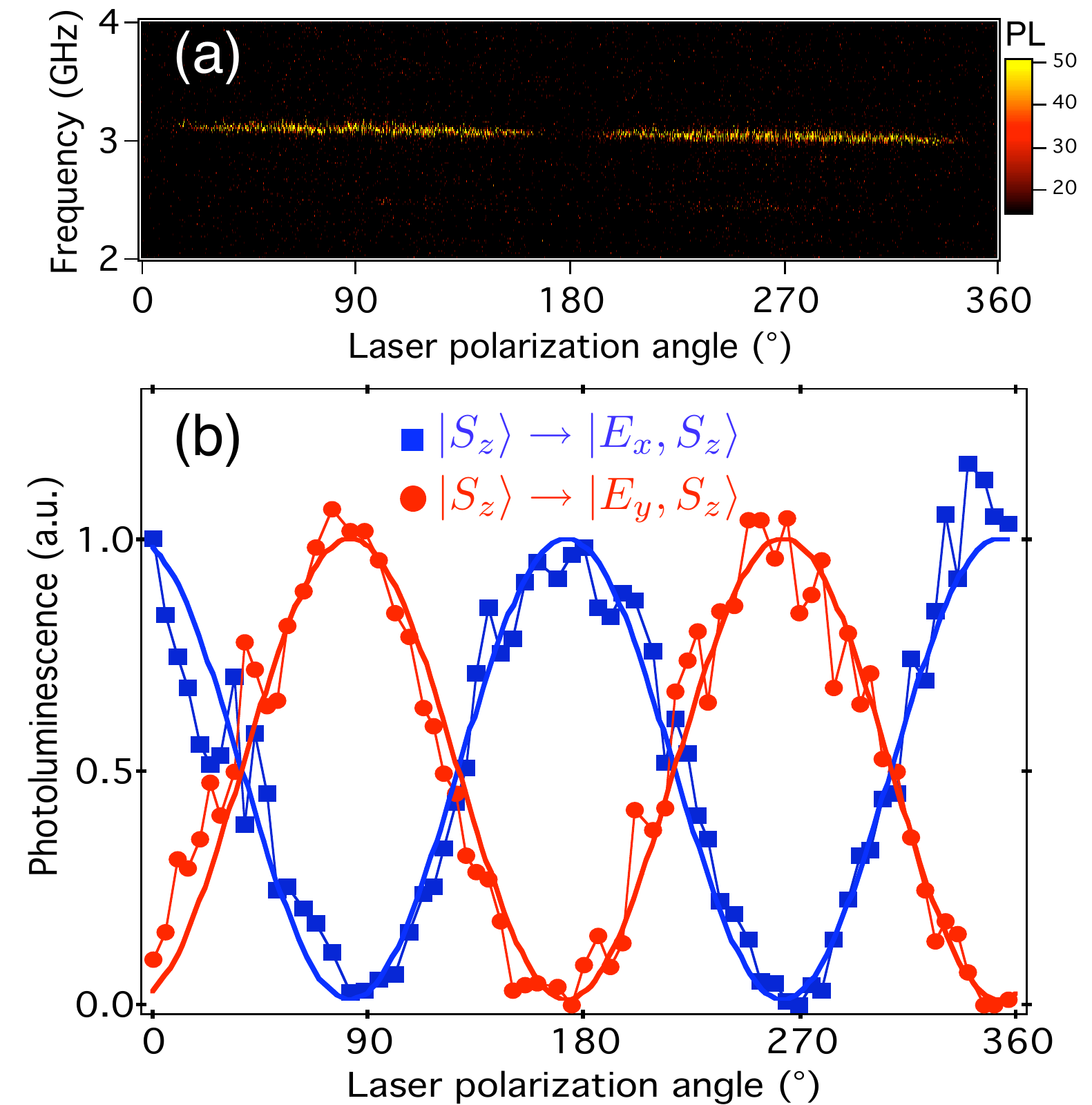}}
\caption{(color online). (a)-Accumulation of excitation spectra recorded by sweeping the laser diode frequency around the optical transition $\left|S_{z} \right\rangle \rightarrow \left|E_{y},S_{z}\right\rangle$ in the lower branch while rotating the laser polarization. Note that the position of the resonance is deflected owing to slow drift of the laser diode frequency. (b)-PL intensity as a function of the laser polarization angle for the optical transition linking $S_{z}$ spin sublevels in the lower branch $E_{y}$ (red) and in the upper branch $E_{x}$ (blue). Solid lines are data fitting using cosine functions.}
\label{Fig2}
\end{figure}

\begin{figure}[t]
\centerline{\includegraphics[width=8.5cm]{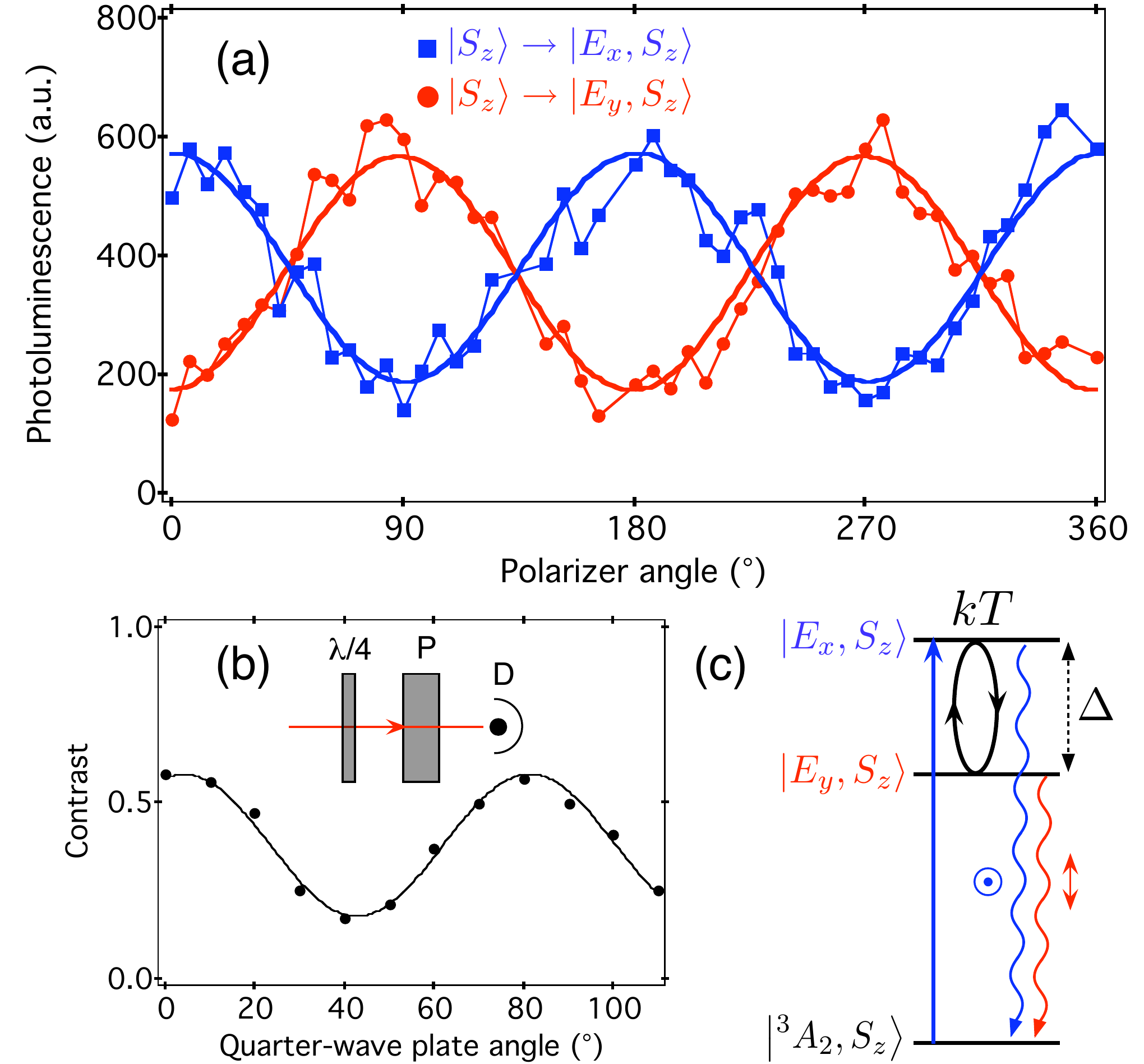}}
\caption{(color online). (a)-PL intensity as a function of the angle of a polarizer (P) installed in front of the single photon detector (D). Red points (resp. blue points) correspond to resonant excitation of the optical transition linking $S_{z}$ spin sublevels in the lower branch $E_{y}$ (resp. in the upper branch $E_{x}$). Solid lines are data fitting using cosine functions. A contrast $C=55\%$ is measured for each excited-state branches. (b)-Contrast of the modulation as a function of the angle of a quarter-wave plate positioned in front of the polarizer. (c)-When the transition $\left|S_{z} \right\rangle \rightarrow \left|E_{x},S_{z}\right\rangle$ is excited resonantly, the thermal bath partially equilibrates the population between the two excited-state branches within the radiative lifetime. Photons are then emitted with different frequencies and orthogonal polarizations.}
\label{Fig3}
\end{figure}

\indent We first investigate the relative orientation of the transition dipoles associated with the two excited-state orbitals $E_{x}$ and $E_{y}$, by rotating the polarization of the laser diode used for optical excitation. Figure 2(a) depicts a typical accumulation of excitation spectra recorded by sweeping the laser frequency around the optical transition $\left|S_{z} \right\rangle \rightarrow \left|E_{y},S_{z}\right\rangle$ in the lower branch $E_{y}$, while rotating the laser polarization. By integrating such excitation spectra, the PL intensity associated with the transition $\left|S_{z} \right\rangle \rightarrow \left|E_{y},S_{z}\right\rangle$ can be displayed as a function of the laser polarization angle (see Fig.~\ref{Fig2}(b)). As expected, a modulation with a contrast closed to unity is observed. The same experiment is then performed by sweeping the laser frequency around the optical transition $\left|S_{z} \right\rangle \rightarrow \left|E_{x},S_{z}\right\rangle$ linking $S_{z}$ spin sublevels in the upper branch $E_{x}$. As depicted in figure~\ref{Fig2}(b), the PL intensity associated with this transition is also modulated, with an opposite phase which is the signature that the two orbital branches are indeed associated to orthogonal transition dipoles in a plane perpendicular to the [111] symmetry axis of the NV defect.\\
\indent We now study the polarization properties of the emitted photons associated with each orbital branch. For such experiments, the laser diode polarization angle is fixed and a polarizer is installed in front of the single photon detector used to detect single NV defect. Following the method described above, excitation spectra are then accumulated by sweeping the laser frequency around a given transition linking $S_{z}$ spin sublevels, while rotating the polarizer in the detection channel. The results of such experiments are shown in figure~\ref{Fig3}. The PL intensity is modulated with an opposite phase depending on which optical transition is selectively excited. However, the contrast of the modulation is measured around $55\%$ for both orbital branches, whereas it should be equal to unity for perfectly polarized emission. Such decrease of the modulation contrast could be explained by a slight elliptical polarization accumulated along the optical path from the sample to the single-photon detector. In order to exclude such possibility, a quarter-wave plate was installed in front of the polarizer in order to compensate any kind of ellipticity introduced in the experimental setup. The contrast of the modulation was then measured for different angle of the quarter-wave plate. As depicted in figure~\ref{Fig3}(b), no improvements could be achieved indicating that the decrease in contrast might result from intrinsic dynamics between the two excited-state orbital branches. Indeed, any relaxation between the two orbitals would decrease the contrast as photons emitted from each orbital branch have orthogonal polarizations.\\
\indent The energy splitting $\Delta$ between the two excited-state orbitals depends on the local strain in the vicinity of the NV defect~\cite{Batalov_PRL2009} and is on the order of $\Delta= 3-10$ GHz in the studied sample. Experiments are performed at $T=4K$, which corresponds to a thermal energy $kT\approx 100$ GHz. Consequently, $kT \gg \Delta$ and the thermal bath can introduce flips between the two orbitals in order to reach the Boltzmann equilibrium. As such a process is not altering the spin projection, it might be fast enough to equilibrate partially the population between the two excited-state orbitals within the radiative lifetime of the NV defect.\\
\indent In the following, we tentatively propose a simple model allowing to derive the relaxation rates between the orbital branches. We denote $p_{x}$ (resp. $p_{y}$) the population of $S_{z}$ spin sublevel in the upper branch $E_{x}$ (resp. in the lower branch $E_{y}$) and we consider excited-state population dynamics before the radiative decay. Neglecting inter-system crossing to the metastable state, the rate equations for this simplified two-level system are given by:
\begin{equation}
 \frac{d}{dt}p_{x}(t)=c_{xy} \, p_{y}(t)-c_{yx}\, p_{x}(t)
\end{equation}
\begin{equation}
p_{x}(t)+p_{y}(t)=1 \,
\end{equation}
where $c_{ij}$ is the transition rate induced by the thermal bath from the orbital branch $E_{j}$ to the orbital branch $E_{i}$. The coefficients $c_{ij}$ fulfill the relation~\cite{DGLR}
\begin{equation}
c_{xy}=c_{yx} \, e^{-\Delta/kT} \ .
\end{equation}
\indent As $kT \gg \Delta$, we then consider $c_{ij}=\gamma$. Assuming for instance that the excitation is resonant with the transition linking $S_{z}$ sublevels in the upper branch $E_{x}$, {\it i.e.} $p_{x}(0)=1$ and $p_{y}(0)=0$, it is straightforward to show that 
\begin{eqnarray}
p_{x}(t)&=&\frac{1}{2}\left(1+ e^{-2\gamma t}\right)\\
p_{y}(t)&=&\frac{1}{2}\left(1- e^{-2\gamma t}\right) \ .
\end{eqnarray}
\indent The average population in each excited-state orbital before radiative decay can then be approximated by $\left\langle p_{x} \right\rangle\approx p_{x}(\tau)$ and $\left\langle p_{y}\right\rangle \approx p_{y}(\tau)$, where $\tau$ is the excited-state radiative lifetime ($\tau\approx 12$ ns). As photons emitted from different excited-state orbital branch have orthogonal polarizations, the PL intensity $I$ is finallly given by
\begin{equation}
I\propto\left\langle p_{x} \right\rangle^{2}\cos^{2}\theta+\left\langle p_{y} \right\rangle^{2}\sin^{2}\theta
\end{equation}
where $\theta$ is the angle between the polarizer and the $x$ axis. We notice that the oscillating term at the beat frequency $\Delta$ is averaging to zero owing to the time response of the photodetector.\\
\indent Rotating the polarizer angle thus leads to a modulation of the PL intensity with a contrast $C$ given by 
\begin{equation}
C=\frac{I_{\rm max}-I_{\rm min}}{I_{\rm max}+I_{\rm min}}=\frac{1-\alpha^{2}}{1+\alpha^{2}}
\end{equation}
where $\alpha=\tanh (\gamma\tau)$. Note that $\alpha=\left\langle p_{y} \right\rangle/\left\langle p_{x} \right\rangle$ when considering a resonant excitation with the upper branch $E_{x}$ whereas $\alpha=\left\langle p_{x} \right\rangle/\left\langle p_{y} \right\rangle$ for a resonant excitation on the lower branch $E_{y}$.\\
\begin{figure}[t]
\centerline{\includegraphics[width=8.5cm]{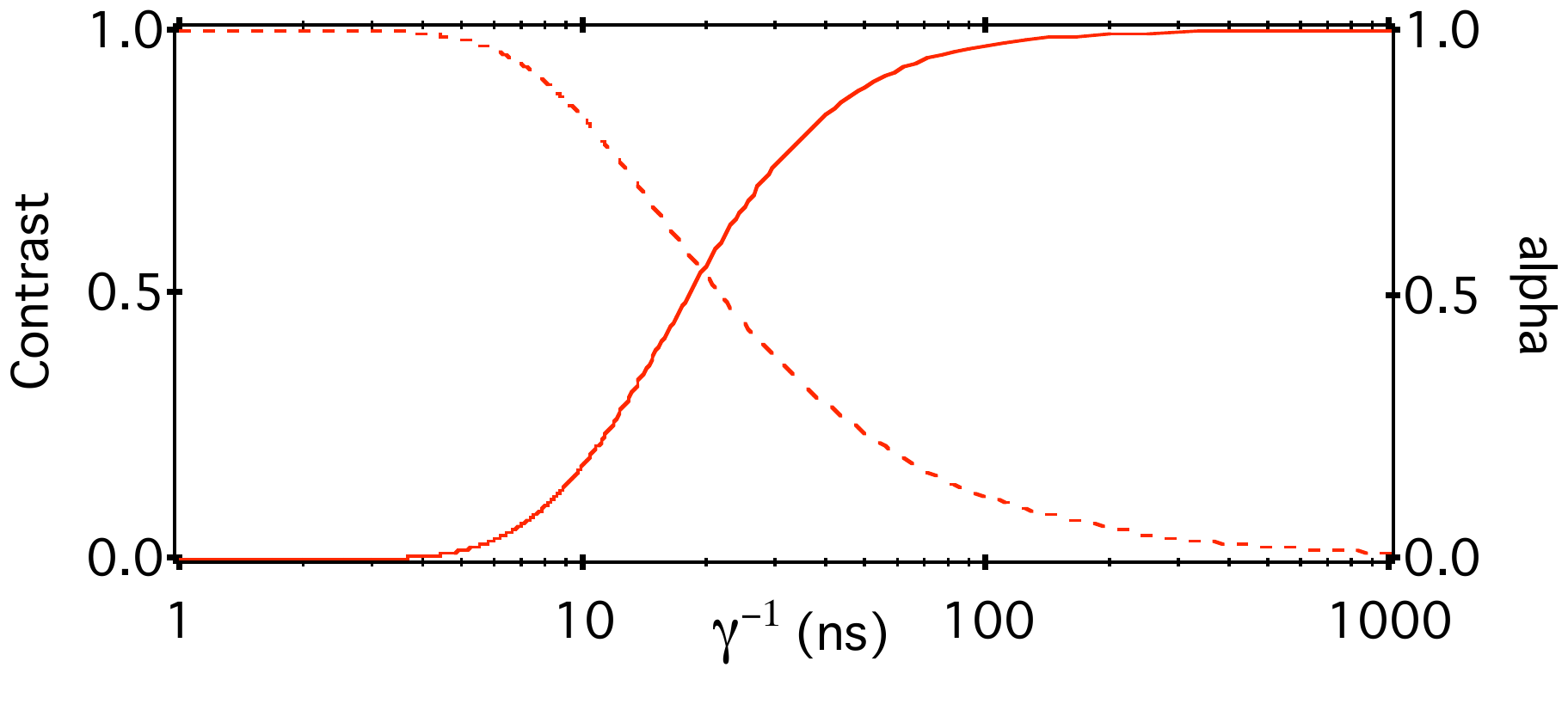}}
\caption{(color online). Contrast of the modulation (solid line) and parameter $\alpha$ (dash line) as a function of $\gamma^{-1}$, using $\tau=12$ ns.}
\label{Fig4}
\end{figure}
\indent If $\gamma^{-1}\ll \tau$, the thermal equilibrium is achieved before radiative decay, leading to $\alpha=1$ and $C=0$. Therefore, the PL is unpolarized. Conversely, if $\gamma^{-1}\gg \tau$, the emission is perfectly polarized as $\alpha=0$ and $C=1$. In intermediate situations, an estimate of the modulation contrast is a direct measurement of the relative population in each of the excited-state orbital branch (see Fig.~\ref{Fig4}). For the data shown in figure~\ref{Fig3}(a), a contrast of $55\%$ is measured, corresponding to $\gamma^{-1}\approx 20 \ \rm ns$ and $\alpha\approx 50 \%$. We notice that modulation contrasts ranging from $0$ to $60\%$ have been measured for different single NV defects. We explain this observations by considering that the transition rate $\gamma$ induced by the thermal bath depends on the local environment of the NV defect and can be fast enough to lead to fully unpolarized emission, {\it i.e.} $\gamma^{-1}< 3 \ \rm ns$ as previously reported in the context of hole-burning studies~\cite{Martin}.\\
\indent In conclusion, using resonant excitation at low temperature, we have shown that NV defect photoluminescence is partially unpolarized. This work gives significant insight into the NV defect dynamics in the excited-state which are important in the context of the realization of indistinguishable single photons using NV defect in diamond. Indeed, owing to fast relaxation between the two excited-state orbitals induced by the thermal bath, indistinguishability could only be obtained by selecting the emission from one excited state orbital branch using a well adjusted polarizer.

\indent The authors are grateful to R.~Kolesov and A.~Nicolet for fruitful discussions. This work is supported by the European Union (QAP, EQUIND, NEDQIT), Deutsche Forschungsgemeinschaft (SFB/TR21) and Landesstiftung Baden-W$\ddot{\rm u}$rttemberg. V. J. acknowledges support by the Humboldt Foundation.

 \end{document}